\documentclass[11pt]{article}
\usepackage{amsmath}
\usepackage{amssymb}
\usepackage{bbm}
\usepackage{epsfig}
\allowdisplaybreaks[4]
\begin{document}
\title{Do LEP results suggest that quarks have integer electric charges?}
\author{P.M. Ferreira \\ CFTC, Universidade de Lisboa, Portugal}
\maketitle
\noindent
{\bf Abstract:} We argue that recent results from two-photon processes at LEP
are better explained by quarks possessing integer electric charges.
\vspace{-9cm}
\vspace{10cm}

Photon-photon collisions have been very thoroughly studied over the past 
years~\cite{rev}. It has been established~\cite{zer,wit} that the photon may
interact both as a point-like particle, the so-called ``direct" processes 
described by perturbative QED and QCD, and as a composite structure, through
``resolved" processes. In the latter it is believed that one of the photons 
first fluctuates into an hadronic state and those partons interact with the 
second photon. Double resolved processes where both photons behave as having a 
structure are also possible. These processes clearly require non-perturbative 
physics and their theoretical description is thus quite complex. Two-photon 
physics has been successful in reproducing most results from particle 
accelerators up to LEP2 energies~\cite{wen} though requiring Reggeon and Pomeron 
parameterizations~\cite{pom} to fit, for instance, the total cross section of 
hadron production by two photons~\cite{opal} - at LEP2 energies, in fact, this 
process is the main source of hadron production. However, recent LEP2 results 
on two-photon physics have raised a problem: both on open heavy quark and 
inclusive hadron production processes, next-to-leading-order (NLO) calculations
in the Standard Model (SM) severely underestimate the data. The SM prediction
for bottom quark pair production via the two photon channel is a factor of 3 
lower than the observed result~\cite{bb}; larger discrepancies are found for the
differential cross section for the inclusive production of $\pi^0$~\cite{L3}, 
$\pi^\pm$ and $K^\pm$~\cite{kch} at high values of the particles' transverse
momentum. Recent reviews may be found in reference~\cite{revi}

A possible explanation for the discrepancy on the $b\bar{b}$ cross 
section was given in ref.~\cite{ber}, requiring a supersymmetric model with a 
light colour octet gluino, although the agreement found there is not 
satisfactory. A hypothesis put forth for the latter discrepancies is 
that of multijet events at high $p_T$ not taken into account in the theoretical 
calculations~\cite{mult}. In this letter we propose a different solution, that 
quarks have integer electric charges and the cross sections under study in 
two-photon processes at LEP2 energies come mostly from perturbative processes. 

Integer charged quark (ICQ) theories were first proposed by Han and 
Nambu~\cite{han}. A renormalisable version was first obtained by Pati and 
Salam~\cite{pati} and several ICQ theories have been obtained via gauge symmetry
breaking, differing on the initial gauge group, generally larger than the SM's. 
Witten remarked~\cite{wit} that the reaction $e^\pm\,e^\pm \rightarrow e^\pm\, 
e^\pm\,q \,\bar{q}$, via the two-photon channel, is the preferential process to 
establish the character of the quarks' electric charges. In fact, as we shall 
shortly see, the perturbative cross sections for this reaction are very 
different for an ICQ theory or a fractionally charged quark (FCQ) one, as
opposed to all reactions involving a single photon: it is a well known fact that
such processes cannot distinguish between ICQ and FCQ models. In this letter we
will compare the FCQ and ICQ predictions for open charm and bottom cross 
sections at LEP2 and show, in section~\ref{sec:open}, that where the SM fails an ICQ theory fits the data very well. In section~\ref{sec:incl} we will 
study inclusive hadron production at LEP2 and again we will show that ICQ 
theories fit the experimental results better than the SM. In 
section~\ref{sec:arg} we will review the experimental evidence for fractional
quark charges and argue it is not conclusive. 

\section{Open heavy quark production in two-photon collisions at LEP2} 
\label{sec:open}

\begin{figure}[htb]
\epsfysize=8cm
\centerline{\epsfbox{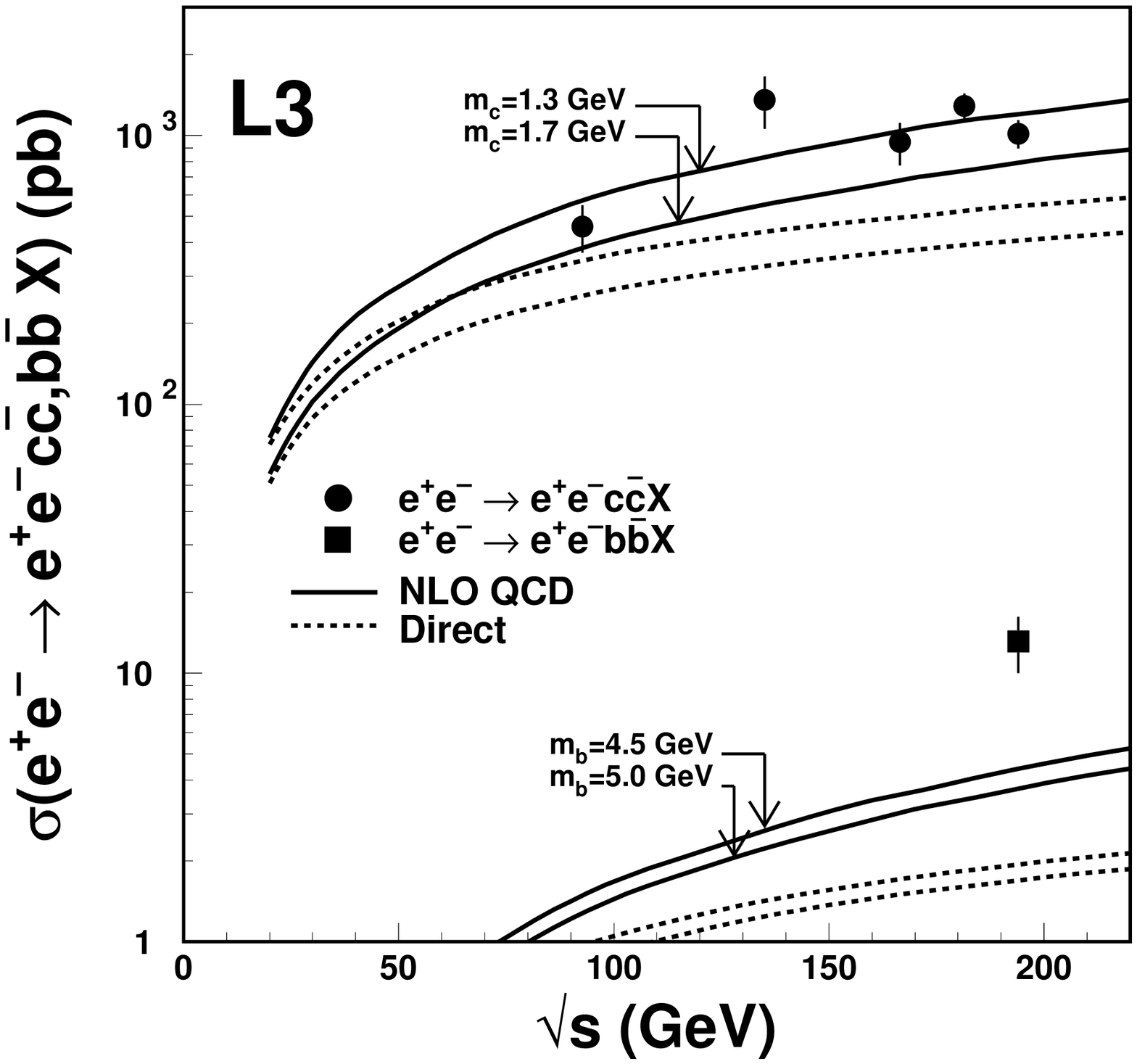}}
\caption{Cross section for production of $c\bar{c}$ and $b\bar{b}$ pairs at
LEP through the two photon channel from the L3 collaboration~\cite{bb}.}
\label{fig:lep}
\end{figure}

As can be appreciated in figure~\eqref{fig:lep}, it is possible to reproduce the
total 
cross section for the production of charm pairs at LEP2 by using contributions 
from both direct and resolved processes, at next-to-leading-order (NLO) in QCD. 
In the same plot we see that even with substantial contributions from resolved 
processes the theory is unable to reproduce the data for bottom quark 
production. The Standard Model prediction is a factor of 3 lower than the 
experimental results. It is clear from fig.~\eqref{fig:lep} that perturbative 
physics alone is insufficient, in the SM, to reproduce the experimental results,
this despite the theoretical predictions having some dependence on the input 
quark mass and the choice of renormalisation scale. Detailed  measurements 
of the photon structure functions revealed the importance of their 
non-perturbative component - at low energies or transverse momentum it is
impossible to disentangle the perturbative contributions from the 
non-perturbative ones. However, it seems a reasonable expectation that at the 
very high LEP2 energies the non-perturbative contributions should be small (that
is certainly true for FCQ theories, see for instance ref.~\cite{dree}). Plus, 
for charm or bottom quark production, the mass of these quarks being so large, 
we have a ``natural" large scale for QCD processes. Once again, it seems 
reasonable to assume that perturbative physics should dominate. We will take 
this as our starting point and calculate only perturbative cross sections.

\begin{figure}[htb]
\epsfysize=8cm
\centerline{\epsfbox{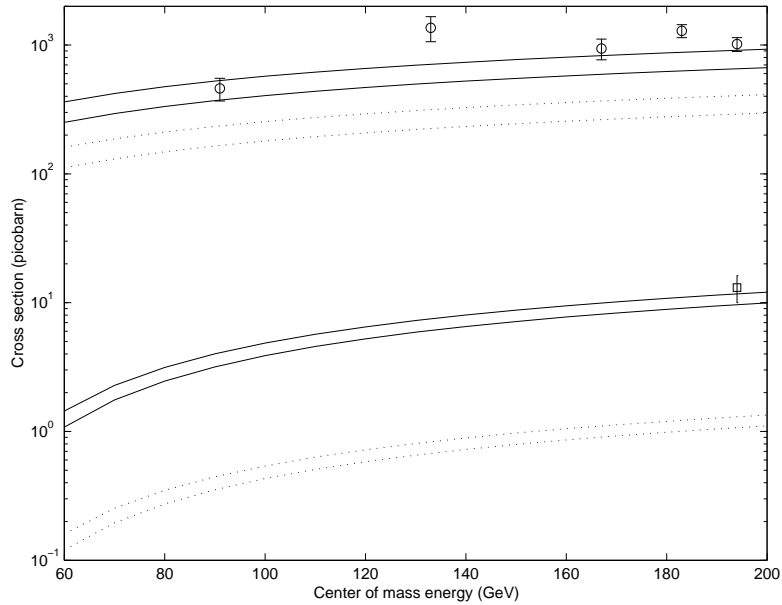}}
\caption{Tree-level cross sections for production of charm and bottom quarks via
the two-photon channel for FCQ (dotted lines) and ICQ (solid lines) models. The
sets of two lines correspond to quark masses varying in the intervals $1.3 \leq 
m_c \leq 1.7$ GeV, $4.5 \leq m_b \leq 5.0$ GeV.}
\label{fig:icq}
\end{figure}

We will see in section~\ref{sec:arg} that single photon reactions give identical
results for ICQ or FCQ theories. There we will give a general argument for why
it is so but for the moment let us consider a particular example, that of the 
quantity $R = \sigma(e^+ e^- \rightarrow q\,\bar{q})/\sigma(e^+ e^- \rightarrow 
\mu^+ \mu^-)$. We begin by noting that we do not observe quarks of a given 
colour individually, rather mesons or baryons containing them; as such the 
amplitude of the process $e^+e^- \,\rightarrow\, q\,\bar{q}$ is the sum of the 
amplitudes for each quark's colour,
\begin{equation}
{\cal M}(e^+e^- \,\rightarrow\, q\,\bar{q}) \;=\; \sum_{i=1}^3 \; {\cal M}(e^+
e^- \, \rightarrow\, q^i\,\bar{q}^i) \;\;\; .
\label{eq:amp}
\end{equation}
Now, the probability of ``finding" a quark of a particular colour in the final
state is $1/3$, so the cross section $\sigma(e^+e^- \,\rightarrow\, q\,\bar{q})$
is proportional to the square of the modulus of the amplitude~\eqref{eq:amp}
multiplied by a factor of $1/3$ and the quantity $R$ is given by
\begin{equation}
R \;=\; \displaystyle{\frac{1}{3}}\,\displaystyle{\sum_q} \,
\left( \displaystyle{\sum_{\mbox{i=colours}}}\, {e_q^i}\right)^2 \;\;\; ,
\end{equation}
where $e_q^i$ is the electric charge of quark $q$ with colour $i$. 
It should be noted that in ICQ theories quarks have charges that depend on
their colour index. An up-type quark, for instance, has charge $(+1, +1, 0)$ for
the quark's colour indices $(1, 2, 3)$. For a down-type quark, the charge is
$(0, 0, -1)$. It is then trivial to see that for both ICQ and FCQ theories we 
obtain the same contributions to $R$ ($4/3$ from an up-type quark, $1/3$ from a 
down-type one). The existing measurements of $R$ are not proof of FCQ - rather,
they make a very good case for the existence of colour. Unlike $R$, the quantity
$R_{\gamma\gamma} = \sigma(e^+ e^-\rightarrow e^+ e^- q \bar{q})/\sigma(e^+ e^- 
\rightarrow e^+ e^- \mu^+ \mu^-)$ (both processes going through the two-photon 
channel) gives very different results for both types of theory, namely (at 
tree-level):
\begin{equation}
\begin{array}{ll}
R_{\gamma\gamma}^{FCQ} \;= & 3\,\displaystyle{\sum_q}\,e_q^4 \\
R_{\gamma\gamma}^{ICQ}\;= & \displaystyle{\frac{1}{3}}\,\displaystyle{\sum_q} \,
\left( \displaystyle{\sum_{\mbox{i=colours}}}\, {e_q^i}^2\right)^2 \;\;\; .
\end{array}
\label{eq:rgg}
\end{equation}
Notice how the ICQ result reduces to the FCQ one when the quark electric charges
do not depend on the colour. From equation~\eqref{eq:rgg} we deduce immediately
that for an up-type quark we would have $R_{\gamma\gamma}^{ICQ}/R_{ 
\gamma\gamma}^{FCQ} \, =\, 9/4$ and for a down-type one, $R_{\gamma\gamma}^{ICQ}
/R_{\gamma\gamma}^{FCQ} \, =\, 9$. Given that the production of muon pairs
occurs identically in both models the cross section for the production of charm
pairs by photon-photon collisions in an ICQ theory is 9/4 the value of the same
quantity calculated in an FCQ model. For bottom production, the ICQ cross
section is 9 times greater than the FCQ one. At this point we are ready to 
compare ICQ and FCQ predictions for open heavy quark production - we follow the 
procedure of ref.~\cite{dree} to calculate $\sigma(e^+ e^- \rightarrow e^+ e^- 
q\bar{q})$ in the Equivalent Photon Approximation~\cite{epa}. We use their 
suggested spectrum of Weizs\"acker-Williams photons and cut-off on the photon's 
maximum virtuality. We set the free quark production thresholds at $(3.8\; 
\mbox{GeV})^2$ (for charm quarks) and $(10.6\; \mbox{GeV})^2$ (for bottom 
quarks). In figure~\eqref{fig:icq} we show the tree-level cross sections. The 
dashed lines are the FCQ results (in agreement with the values of 
ref.~\cite{dree}, the solid lines the ICQ ones. The comparison with 
fig.~\eqref{fig:lep} is revealing - the ICQ theory not only describes as well 
the production of charm pairs, it also succeeds where the SM fails, the 
production of bottom quarks. 

The QCD corrections to this calculation should improve the agreement, as we 
expect them to increase the tree-level cross section (by no more than 30\% for 
the SM). In ICQ theories gluons generally acquire mass and electric charge; the 
QCD corrections will depend on the particular model we choose. We give two 
examples: in ref.~\cite{god} a model was considered with eight massive gluons of
mass $\sim 0.3$ GeV to explain the $\gamma\gamma$ results from PETRA. More 
recently massive gluons of mass about 1 GeV were considered to explain the 
radiative decays of the $J/\Psi$ and $\Upsilon$ mesons~\cite{fiel}. For such 
reasonably low gluon masses we expect the NLO QCD corrections from the SM to be 
similar in form to those of the ICQ theory. A good estimate of the NLO QCD ICQ 
cross section is therefore to multiply the ``direct" results of 
fig.~\eqref{fig:lep} by factors of 9/4 and 9 for the charm and bottom quarks 
respectively. As can be seen from table 1 the agreement between data and these 
estimated ICQ predictions is, again, excellent. 
\begin{table}[t]
\begin{center}
\begin{tabular}{|c|c|c|} \hline & & \\
 $\sqrt{s}$ (GeV) & ICQ predictions & Exp. Data  \\ & & \\ \hline & & \\
 91 & 563 - 722 & $460 \pm 92.2 $ \\ & & \\ \hline & & \\
 133 & 680 - 925 & $1360 \pm 300$ \\ & & \\ \hline & & \\
 167 & 770 - 982 & $940 \pm 172.1$ \\ & & \\ \hline & & \\
 183 & 817 - 1125 & $1290 \pm 149.7$ \\ & & \\ \hline & & \\
 194 & 900 - 1181.3 & $1016 \pm 123.7$ \\ & & \\ \hline & & \\
 194 ($b\bar{b}$) & 14 - 16 & $13.1 \pm 3.1$ \\ & & \\ \hline
\end{tabular}
\caption{Comparison of estimated ICQ NLO QCD predictions with experimental
data~\cite{bb}. The ICQ cross sections are obtained from the ``direct" curves of
fig.~\eqref{fig:lep}. All cross sections are in picobarn and, except for the last 
line, refer to the reaction $e^+ e^- \rightarrow e^+ e^-\, c\,\bar{c}$.}
\end{center}
\end{table}
Recently~\cite{eu} an ICQ model with three massless gluons and five massive
ones, with masses larger than $\sim 76$ GeV, was proposed. In such a model the
gluon bremsstrahlung corrections will involve only three gluons, not eight. 
Also, given that the gluons' mass is so large, an approximation to the one-loop
cross section would be to consider the contributions of the massless gluons
alone. So, from the results of ref.~\cite{dree}, an estimate of the 
one-loop QCD correction to the FCQ cross section would be
\begin{equation}
\sigma(\gamma\gamma \rightarrow q\,\bar{q})_{1-loop} \; \simeq \;  
\frac{12\pi\alpha^2 e_Q^4}{s_{\gamma\gamma}}\,\beta\,\frac{3\alpha_s}{4\pi}\,
\left[\frac{\pi^2}{2\beta}\,-\, \left(5 - \frac{\pi^2}{4}\right) \,+\, O(\beta) 
\right]\;\;\; ,
\end{equation}
where $s_{\gamma\gamma}$ is the two-photon total energy squared and $\beta = 
(1 - 4\,m_q^2/s_{\gamma\gamma})^{1/2}$, with $m_q$ the quark mass. With three
massless gluons the $SU(2)$ Casimir $3/4$ appears. Repeating the calculations 
that led to fig.~\eqref{fig:icq} we see that this estimated correction 
increases the cross sections by no more than $\sim 7\%$. This is probably an 
underestimation of one-loops effects and it improves the agreement with
the experimental results only slightly.

An exact calculation for particular ICQ models, including NLO QCD corrections, 
is clearly necessary at this stage. Such calculation could put bounds on the 
masses of the ICQ gluons (or the colour-breaking vevs of ref.~\cite{eu}). As 
striking as the numerical agreement is the fact that we obtained these results 
from a purely perturbative ICQ theory. More detailed calculations or 
measurements would indicate if there is the need of a (small, surely) 
non-perturbative component to the cross sections at these high energies. A small
resolved component seems necessary to improve the charm cross section, judging 
by our estimated results. Let us also emphasize that the analysis that led to 
the experimental data presented in refs.~\cite{bb} assumed normal QCD 
backgrounds. Rigorously that analysis would have to be re-made assuming 
a gluon (massive or otherwise) background from an ICQ theory, but it seems 
reasonable to expect the end results would not be significantly different from 
those now available. 

\section{Inclusive hadron production in two-photon collisions at LEP2}
\label{sec:incl}

Unlike the total $c\bar{c}$, $b\bar{b}$ cross section, even a leading order (LO)
calculation  for an inclusive process requires some non-perturbative input in
the form of fragmentation functions. In fact, in inclusive processes we are
interested, not in the result of the partonic interaction (like in the open 
channels) but rather on a specific hadron appearing from the hadronisation of 
quarks. For the reaction $e^-(P_1)\,+\,e^+(P_2)\,\rightarrow h\,(P_h)\,+\,X$ via
the two-photon channel the differential cross section for inclusive production 
of a given hadron $h$ (with four-momentum $P_h$, rapidity $y$ and transverse 
momentum $p_T$ is given, at LO, by
%%%
\begin{equation} 
\frac{\partial \sigma}{\partial y\,\partial p_T}= \frac{2p_T}{S} \sum_l \,
\int_{1-V+VW}^{1}\,\frac{dz}{z^2}\,\int_{VW/z}^{1-(1-V)/z}
\frac{dv}{v\,(1-v)}\,f_-(x_1)\,f_+(x_2)\, D_l^h(z,M_F^2)\,\frac{d
\sigma_{\gamma\gamma\rightarrow l\bar{l}}}{dv} \;\,
\label{eq:dif}
\end{equation}
where $S$, $V$ and $W$ are the usual Mandelstam variables, defined by $S = (P_1
+P_2)^2$, $V = 1 + T/S$ and $W = -U/(S+T)$, with $T = (P_1 - P_h)^2$, $U = (P_2
-Ph)^2$. We also define the partonic Mandelstam variables $s$, $t$, $u$ and $v$
using the four-momenta of the partons instead of that of the external particles:
if $P_1^\gamma$, $P_2^\gamma$ and $P_q$ are the momenta of the photon emitted by
the electron and positron and the momentum of the hadronising quark we have 
$s = (P_1^\gamma+P_2^\gamma)^2$, $t = (P_1^\gamma - P_q)^2$ $u=(P_2^\gamma-P_q
)^2$ and $v = 1 + t/s$. $s$ is clearly the center-of-mass energy of the 
two-photon system. In equation~\eqref{eq:dif} $x_1$, $x_2$ and $z$ are the 
momentum fractions carried by both photons and the hadron (thus defined as 
$P_1^\gamma = 
x_1 P_1$, $P_2^\gamma = x_2 P_2$ and $P_h = z P_q$), related to each other by
$x_1 = VW/vz$ and $x_2 = (1-V)/z(1-v)$. $f_-(x_1)$ and $f_+(x_2)$ are the 
probability functions for finding a photon inside the electron or positron with 
momentum fractions $x_1$ and $x_2$ respectively. $D_l^h(z\,,\,M_F^2)$ is the 
fragmentation function of a quark of flavour $l$ into a hadron $h$ carrying a 
fraction $z$ of $l$'s momentum. It is evaluated at an {\em a priori} unknown
fragmentation scale $M_F$. Finally, the last term in this expression is given
by~\cite{dree, kra}
%%%%
\begin{equation}
\frac{d \sigma_{\gamma\gamma\rightarrow l\bar{l}}}{dv} \; = \; \frac{2\pi
\alpha^2\,C_F}{s}\; \left[\frac{t}{u} \, + \, \frac{u}{t}\, + \, 4\,\frac{s\,
m^2}{t\,u} \,-4\,\left(\frac{s\,m^2}{t\,u}\right)^2  \right]\;\;\;, 
\end{equation}
where $m$ is the mass of the quark $l$ (set to zero if $l\,=\,u\,,\,d\,,\,s$) 
and $C_F$ is a colour factor, related to the fourth power of the quarks'
electric charges. For an FCQ theory, $C_F$ equals $16/27$ and $1/27$ for up and 
down type quarks respectively. For an ICQ theory, those factors become $4/3$
and $1/3$. The fragmentation functions $D_h^l$ are not known from first 
principles: they must be extracted from experimental data and are inherently 
non-perturbative in nature (though their evolution with the scale $M_F$ is 
governed by an Altarelli-Parisi type equation). For pions, for instance, NLO 
parameterizations were first given in~\cite{greco}. The current state-of-the-art 
fragmentation functions, parameterized to include scale dependence in a very 
useful manner, were elaborated by Kniehl {\em et al}~\cite{kni} and will be used
throughout this paper (at LO only for consistency, since we use the tree-level 
partonic cross section in eq.~\eqref{eq:dif}).

\begin{figure}[htb]
\epsfysize=8cm
\centerline{\epsfbox{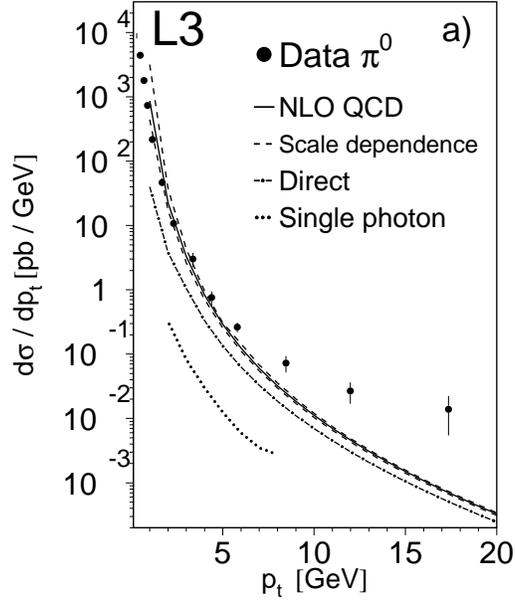}}
\caption{Inclusive differential cross section for $\pi^0$ production in 
photon-photon collisions at LEP2, for $|\eta|<0.5$ and $s > 5$ GeV. The data 
is compared to a 
NLO QCD calculation, with the direct contribution represented by the dot-dashed
line. L3 collaboration~\cite{L3}. }
\label{fig:l3pi0}
\end{figure}

At this point let us consider the current experimental data from the L3 
collaboration for inclusive $\pi^0$ photoproduction at LEP2~\cite{L3}. As is 
plain from fig.~\eqref{fig:l3pi0} there is a huge discrepancy between the NLO 
QCD SM prediction and the actual data for $p_T \geq 5$ GeV, a discrepancy that 
grows worse for higher values of $p_T$: for $p_T \simeq 17$ GeV the SM 
prediction is roughly 6 times smaller than the experimental data. This is not a 
new fact: Gordon~\cite{gor} compared the first NLO QCD calculation for 
inclusive $\pi^0$ photoproduction with MARKII data~\cite{mult} and concluded
that a discrepancy between theory and experiment at high $p_T$ was already
present. The other important piece of information to retain from 
fig.~\eqref{fig:l3pi0} is that at low $p_T$, as expected, it is the 
non-perturbative resolved contributions that dominate, whereas at high $p_T$ the
direct processes constitute the bulk of the cross section. We will perform only 
a LO perturbative calculation so we cannot expect to find good agreement in the
low $p_T$ region - but an ICQ theory should increase the high $p_T$ agreement
considerably. Consider: in an ICQ theory the partonic cross section 
$\sigma_{\gamma\gamma \rightarrow q\bar{q}}$ in eq.~\eqref{eq:dif} is multiplied
by a factor of 9/4 ({\em vis a vis} the FCQ cross section) for q = u, c and a 
factor of 9 if q = d, s, b. However, unlike the open channel, we do not expect
the ICQ cross section to be given simply by an overall multiplicative constant
times the FCQ result - in fact, the charm quark contribution to 
eq.~\eqref{eq:dif} does not ``kick in" until $s$ has values superior to the
charm production threshold (set to 2.97 GeV in the fragmentation functions of
ref.~\cite{kni}). Likewise the bottom contribution only affects the cross
section for $s$ larger than the $b\bar{b}$ threshold, set to 9.46 GeV. Because
$s = p_T^2/z^2\,v\,(1-v)$ the ICQ contributions are therefore larger, compared
to the FCQ ones, for larger $p_T$: for instance, if above the charm threshold 
the FCQ charm contribution is, say, $X$, the ICQ term will be $9\,X/4$. Above 
the $b\bar{b}$ threshold the bottom contribution would be $Y$, so the FCQ cross
section would be given by $X + Y$; the ICQ one would thus equal $9\,X/4 \,+\,
9 Y$. So, the ICQ cross section is always larger than the FCQ one, and the 
ratio between the two is expected to grow with $p_T$. This behaviour is exactly 
what we need to explain the discrepancy in fig.~\eqref{fig:l3pi0} - multiplying
the direct FCQ cross section by a single numeric factor would be useless, but 
the ICQ prediction is that that factor increases with the value of $p_T$. 

\begin{figure}[htb]
\epsfysize=8cm
\centerline{\epsfbox{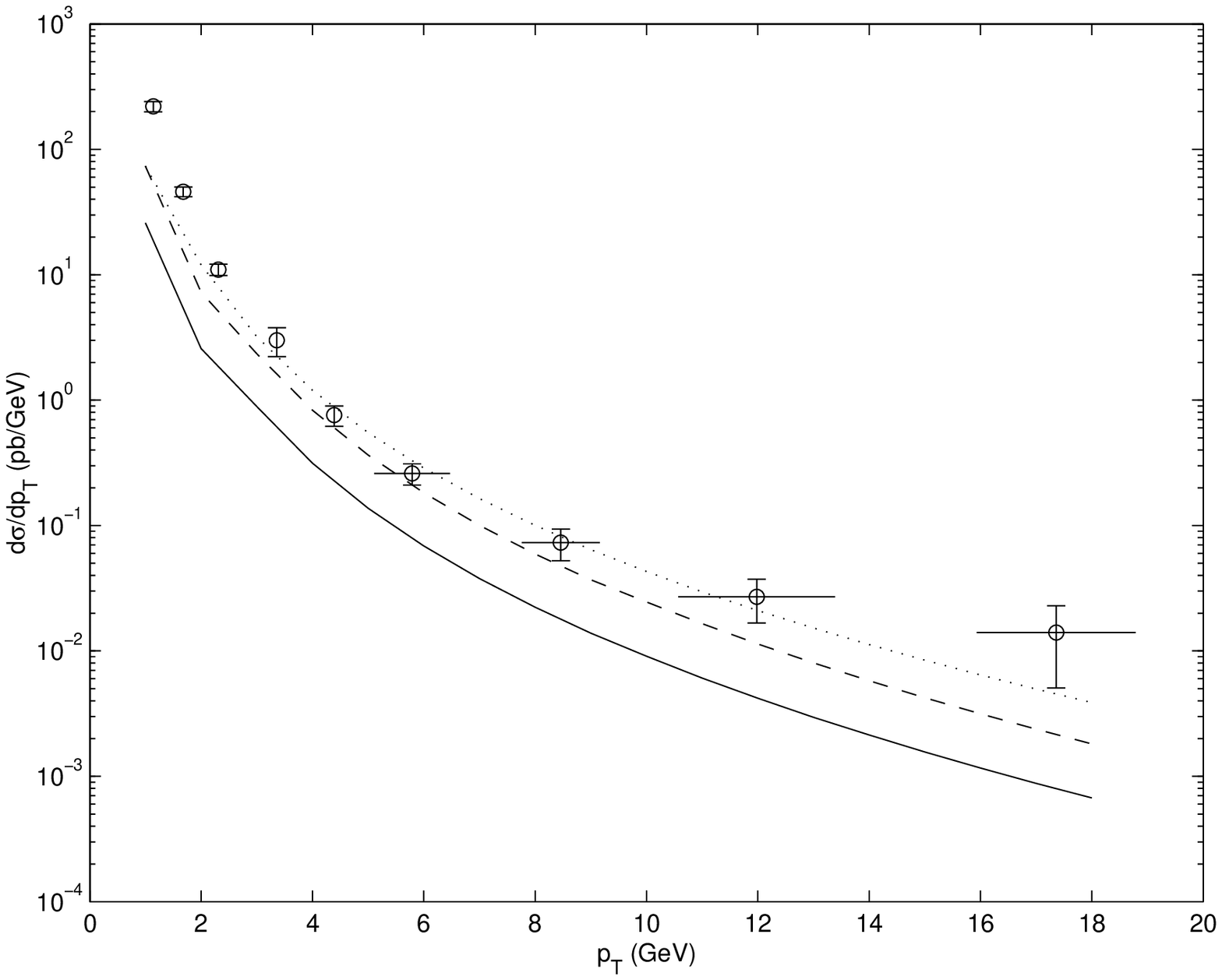}}
\caption{Inclusive differential cross section for $\pi^0$ production in
photon-photon collisions at LEP2, for $|\eta|<0.5$ and $s > 5$ GeV - the data 
is compared to a
tree-level SM calculation (solid line) and ICQ calculations with $M_F = p_T$
(dashed line) and the lowest value of $M_F$ possible, described in the text 
(dotted line).}
\label{fig:pi0}
\end{figure}

We follow closely the calculation of ref.~\cite{gor}, adopting their
Weizs\"acker-Williams~\cite{epa} spectrum; the LO fragmentation functions
of~\cite{kni} are used with the fragmentation scale initially set to the
standard value $M_F \,=\,p_T$. The tree-level cross section we present in
figure~\eqref{fig:pi0} is in agreement with the results of~\cite{gor,L3}.
Multiplying the partonic cross sections by the appropriate factors gives us the
ICQ cross section - it is clear there is a considerable improvement over
the FCQ prediction for high $p_T$. Obviously, at low $p_T$ we do not agree (FCQ
or ICQ) with the data. As explained that was to be expected, since in that
region the resolved processes are dominant and we are not including them in this
calculation. The horizontal error bars in this plot are calculated from the 
power law fits to the cross section performed by the L3 collaboration: for $p_T 
\geq 1.5$ GeV they found that the cross section goes like $\sim 
p_T^{-4.1}$~\cite{L3}. With this power law it is a trivial matter to calculate
the root mean square deviation for the $p_T$ bin intervals considered in the
experiments. For instance, for the 15-20 GeV bin interval the mean value found
is 17.36 GeV and the error bar we find is 1.4 GeV, which seems quite
reasonable~\footnote{I thank Pablo Achard for clarifying this issue for me.}. 
If, like Gordon~\cite{gor}, we push the fragmentation functions to their limit 
and evaluate each quark's contribution at the lowest fragmentation scale allowed
by the parameterizations of~\cite{kni} ($\sqrt{2}$ GeV for the light flavours, 
the quark's mass for the two heavy ones), we find, as he did, an increase in the
cross section. The agreement with the data at high-$p_T$ is then found to be
very good. This may seem an {\em outr\'e} choice of fragmentation scale but it
must be said it is as justified, in principle, as taking $M_F = p_T$~\footnote{
Lest the reader thinks we are dishonestly favoring ICQ theories let them be
assured that even considering the $p_T$ error bars and the same extreme choice 
of fragmentation scale the discrepancy between the data and the SM predictions 
remains immense.}. This choice of scale, in fact, allows us to reproduce the 
data for $p_T$ as low as $\sim $ 3 GeV without the need for any resolved 
contributions.

\begin{figure}[htb]
\epsfysize=8cm
\centerline{\epsfbox{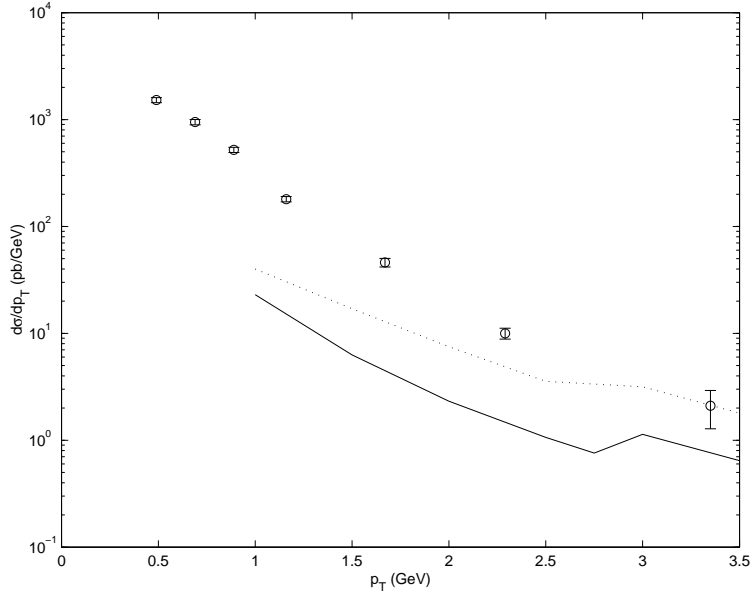}}
\caption{Inclusive differential cross section for $K^0_S$ production in
photon-photon collisions at LEP2, for $|\eta|<1.5$ and $s > 5$ GeV  - the data 
is compared to a
tree-level SM calculation (solid line) and ICQ calculation with $M_F = p_T$
(dotted line). The structure at $p_T \simeq 3$ GeV is due to the charm 
production threshold in the fragmentation functions~\cite{kni}.} 
\label{fig:k0}
\end{figure}

\begin{figure}[htb]
\epsfysize=8cm
\centerline{\epsfbox{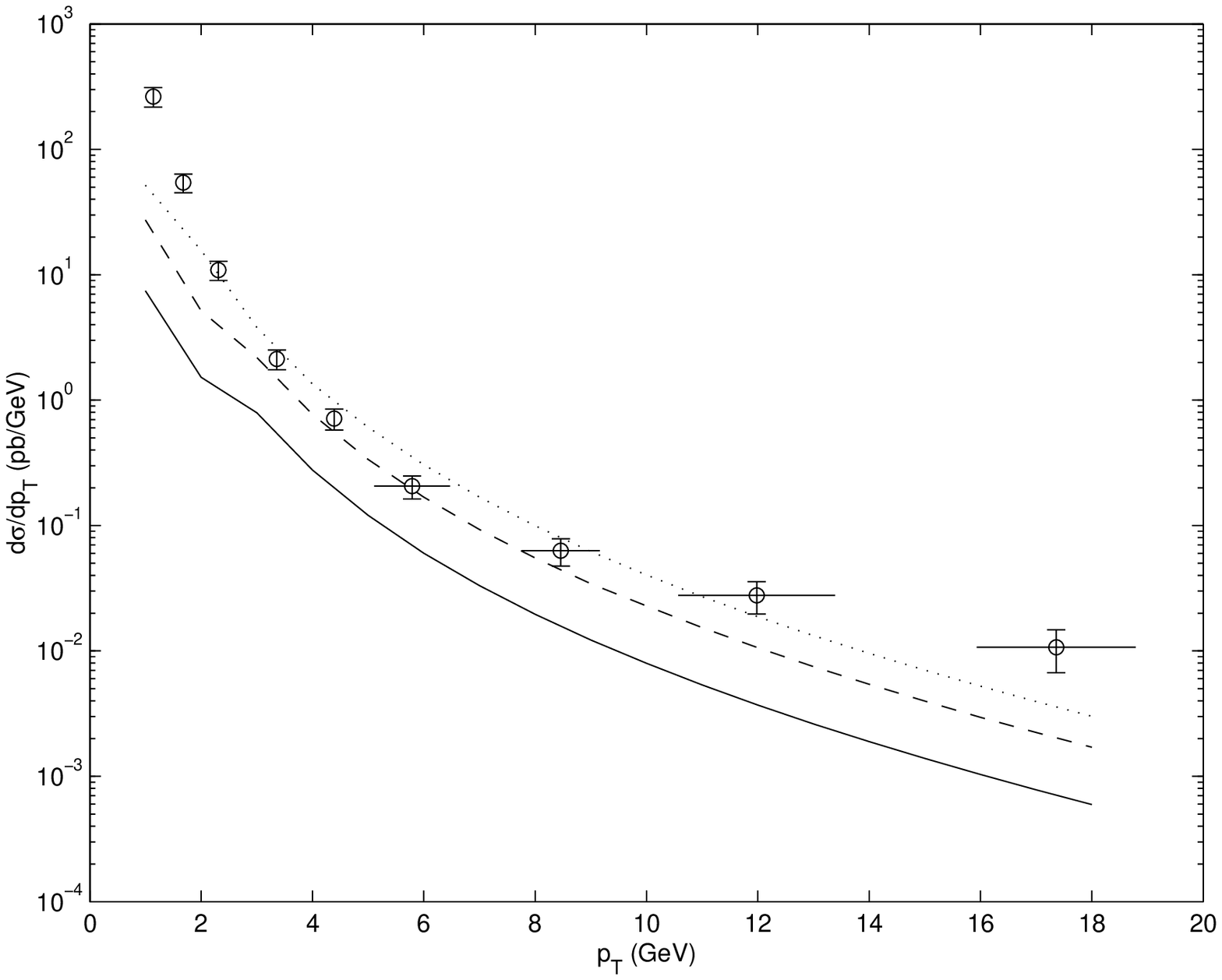}}
\caption{Inclusive differential cross section for $K^\pm$ production in
photon-photon collisions at LEP2, for $|\eta|<1$ and $s > 5$ GeV - the data 
is compared to a
tree-level SM calculation (solid line) and ICQ calculations with $M_F = p_T$
(dashed line) and the lowest value of $M_F$ possible, described in the text
(dotted line).}
\label{fig:kch}
\end{figure}

\begin{figure}[htb]
\epsfysize=8cm
\centerline{\epsfbox{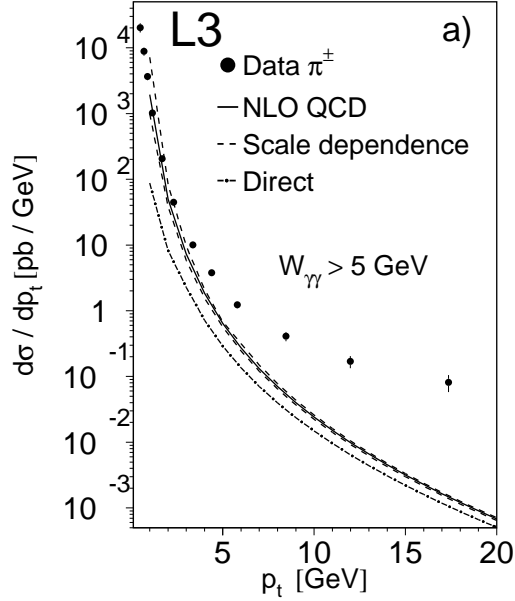}}
\caption{Inclusive differential cross section for $\pi^\pm$ production in
photon-photon collisions at LEP2, for $|\eta|<1$. The data are compared to a
NLO QCD calculation, with the direct contribution represented by the dot-dashed
line. L3 collaboration~\cite{kch}. }
\label{fig:l3pich}
\end{figure}

\begin{figure}[htb]
\epsfysize=8cm
\centerline{\epsfbox{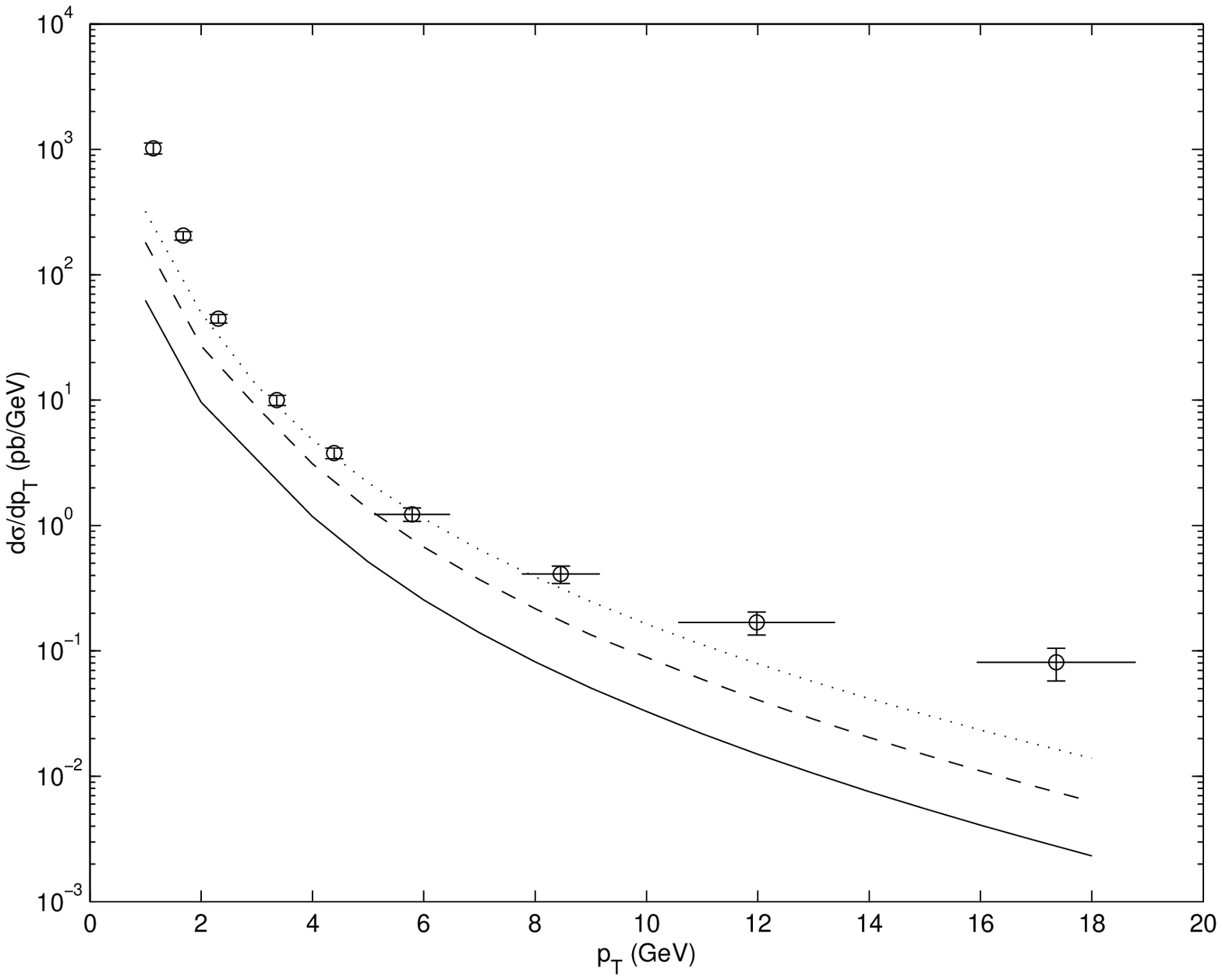}}
\caption{Inclusive differential cross section for $\pi^\pm$ production in
photon-photon collisions at LEP2, for $|\eta|<1$ and $s > 5$ GeV - the data 
is compared to a
tree-level SM calculation (solid line) and ICQ calculations with $M_F = p_T$
(dashed line) and the lowest value of $M_F$ possible, described in the text
(dotted line).}
\label{fig:pich}
\end{figure}

We next consider the $K_S^0$ results from~\cite{L3} - these are perfectly
reproduced by the SM predictions (see figure 3.b from ref.~\cite{L3}) but we 
remark they do not extend to large values of $p_T$. It would be interesting to 
check if, for large $p_T$, one finds a discrepancy like that seen in the $\pi^0$
cross section. For such low values of $p_T$ we cannot expect to fully reproduce the 
$K_S^0$ data but we must confirm if the ICQ hypothesis does not ruin a possible 
agreement with experiment (in other words, we need to verify that the ICQ 
predictions are not larger than the data). As we see in fig.~\eqref{fig:k0}, 
again the low-$p_T$ points are not reproduced by the ICQ curve but 
interestingly, the experimental point immediately above the charm threshold is -
this agrees nicely with our hypothesis in the previous section, in which we 
considered that charm production was a process already dominated by perturbative
contributions. Recent results for inclusive charged hadron production in
two-photon collisions~\cite{kch} have also revealed large discrepancies between 
the SM predictions and the experimental results. For charged kaons our results 
are shown in fig.~\eqref{fig:kch}. Again, we see that the ICQ calculation 
with the standard choice $M_F = p_T$
improves the agreement with the experimental results; the same choice of lowest 
possible fragmentation scale gives very good agreement for all values of $p_T$ 
larger than about 2 GeV, except for the final point~\footnote{The error bars 
shown in fig.~\eqref{fig:kch} were calculated in the same manner as those of 
fig.~\eqref{fig:pi0}.}. The deviation there is small and would surely be 
overcome if NLO contributions were taken into account. More problematic
are the charged pion results, as this is the channel for which the deviation
between SM prediction and experimental result is largest - a factor of almost
40, judging from fig.~\eqref{fig:l3pich}. We see from fig.~\eqref{fig:pich} that
with the ``normal" choice of fragmentation scale the ICQ theory only fits the
data until $p_T \simeq 6$ GeV. Pushing the fragmentation functions to their
limit we get agreement up to about 12 GeV, but a large discrepancy remains for
the experimental point with the highest value of $p_T$, a
factor of about 3. One could say that it is impressive enough that with a 
tree-level calculation alone we passed from a factor of 40 to a factor of 3, but
the disagreement with the data remains.

A general comment about these results: they are strongly dependent on the 
fragmentation functions, as we showed by varying the fragmentation scale. The
improvement over the FCQ predictions, regardless of that dependence, is 
unequivocal, but good agreement with experimental results is achieved in
three cases only with a peculiar (if allowed) choice of scale. The disagreement 
we found for the case of charged pions might have to do with a feature observed 
by the L3 collaboration - they compared, in ref.~\cite{kch}, the $\pi^\pm$ data 
obtained with the $\pi^0$ results from~\cite{L3}; it was expected that the ratio
between both sets of data points should be around 4, a factor of 2 due to 
doubled $\eta$ coverage for charged pions - an assumption verified by the data, 
see figure~(3) of~\cite{kch} - and a factor of 2 from the assumption in the 
fragmentation functions of~\cite{kni} that $D_q^{\pi^0}(z\,,\,M_F^2)\,=\,
D_q^{\pi^\pm}(z\,,\,M_F^2)/2$. Now, this latter assumption is well motivated by 
various sets of experimental data (see~\cite{kni} and references therein) but 
the fact remains that (see figure~(2.b) of~\cite{kch}) for $p_T \geq 4$ GeV the 
ratio between $\pi^\pm$ and $\pi^0$ data seems to be centered around 6, not the
expected value of 4. So perhaps the data is showing us that the assumption
relating the fragmentation functions of neutral and charged pions is incorrect
in this domain of $p_T$ and $s$. The question remains open. Another important 
remark: these are LO results. The NLO fragmentation functions are not largely 
different from the LO ones, but clearly a NLO calculation in specific ICQ models
is necessary. Finally, agreement with the high-$p_T$ data is found but the 
question remains as to whether one hasn't destroyed the agreement found for 
small values of transverse momentum within the SM, where non-perturbative
calculations dominate. The obvious comment to make is that the resolved 
contributions considered in refs.~\cite{L3,kch} were calculated in the context 
of the
SM; to verify whether an ICQ model agrees with the data for low $p_T$ it would
be necessary to repeat those calculations within the framework of the 
particular model under study.  

\section{Evidence for fractional quark charges}
\label{sec:arg}

The SM is an extraordinarily successful theory, a fact reinforced by precision 
measurements agreeing with two and three loop calculations made in the framework
of the Glashow-Weinberg-Salam theory. And yet, as we shall shortly see, direct 
experimental evidence for fractional electric quark charges is very limited. 
This is not surprising: due to confinement we have no direct access to quarks 
and must study the results of their hadronisation, a task complicated by 
non-perturbativity at low energies. An overview of the experimental 
evidence for the nature of quarks' charges was done elsewhere~\cite{rind}, we 
analyze in detail some of the evidence for FCQ usually quoted in the literature 
and try to argue that ICQ models are not excluded by it. Before we start we 
point out that in what concerns purely leptonic processes both models are 
obviously identical. 

\begin{flushleft} $\bullet$ {\bf Single photon processes:} 
as was mentioned in section~\ref{sec:open} the
$R$ quantity is usually presented as evidence for FCQ; and yet we showed it is
predicted identically for both FCQ and ICQ theories. This is an example of an
interesting feature of ICQ theories - for any
process involving a single photon they mimic FCQ models (at tree level at 
least). This can be shown in a general manner if we recall that in ICQ theories,
due to different gauge symmetry breaking, the electromagnetic current $J_\mu^I$
is the sum of two pieces: $J_\mu^I\,=\,J_\mu^F\,+\,J_\mu^8$ where $J_\mu^F$ is 
the usual FCQ current, corresponding to the electric charge operator $Q_F\,=\,
T_3\,+ \,Y/2$. $J_\mu^8$ is the extra ICQ contribution, non-zero only for 
particles
with colour - the charge operator associated with it is $Q_8\,=\,\lambda_8/
\sqrt{3}\,=\,\frac{1}{3}\mbox{diag}(1\,,\,1\,,\,-2)$, where $\lambda_8$ is the
eighth Gell-Mann matrix. As a consequence in an ICQ theory the electric
charge of a quark becomes colour dependent, for an up-type quark it is $q_u =
(+1\,,\,+1\,,\,0)$ and for a down-type one, $q_d =(0\,,\,0\,,\,-1)$. The 
fundamental thing to retain from this is the fact the ICQ current is the sum of
the FCQ one with a term that is traceless in colour space. Now, we know colour 
is not observed in nature, only $SU(3)$ singlets. If the initial and final 
states of any given process are colour singlets then the basic tenets of quantum
theory tell us that we must sum the amplitude on the colour indices and divide 
its square by the total number of colours (as if we were dealing with an 
interference experiment). The sum on the colours will correspond to taking 
the amplitude's trace in colour space. Thus, for any process involving a single 
photon (a single current, then), the extra ICQ contribution is cancelled and the
result is the same of an FCQ theory. The current $J_\mu^I$ being the result of
gauge symmetry breaking, this result holds for any order of perturbation theory
as ICQ theories are renormalisable. 
\end{flushleft}

ICQ and FCQ theories are therefore indistinguishable for single photon processes
(also single $Z^0$ processes - see ref.~\cite{eu}, for instance). This would in
principle include deep inelastic scattering experiments (two-photon interactions
in these are an extremely small correction to the main interaction) although 
in studying these one will have to take into account the different spectrum of 
gluons of ICQ models. ICQ theories also predict identical rates for (most) meson
radiative decays. There is a wealth of data on radiative decays of heavy mesons
(for reviews, see for instance~\cite{nov,bes}), but these involve dipolar 
transitions between different states of the quark bound states, with the
emission of a single photon. And because a single photon is involved both ICQ
and FCQ theories would have the same predictions for those processes. Of course,
cascade decays with the emission of several photons are a succession of single
photon decays and as such are identical in both models. 

The only way to obtain a difference between ICQ and FCQ models in photonic 
processes is to consider a situation where at least two $J^I$ currents are
involved; this will lead to a transition amplitude between two states $|i>$ and
$|f>$ of the form $<i|\ldots\,J^I_\mu\,J^I_\nu\,\ldots|f>$. The end result must 
still be a colour average but this time there will be a $(J^8)^2$ term which is 
not cancelled and produces a substantial difference between ICQ and FCQ, as was
apparent in the values of $R_{\gamma\gamma}$ found earlier. 

\begin{flushleft} $\bullet$ {\bf Jet charges:} Feynman and 
others~\cite{jet} proposed a clever way
of measuring the quarks' electric charges using neutrino-nucleon inelastic
scattering. In a $\nu -p$ collision, if the final state includes a positron
then an interaction with the exchange of a $W^+$ boson surely occurred and 
inside the proton only $d$ or $s$ quarks could interact with the $W^+$. Thus the
jets produced in this reaction will almost certainly result from the 
hadronisation of an
up-type quark and a measurement of their average charges should yield 
information about the quark's own. The experiments were performed~\cite{jete} 
and the results agree with fractional quark charges. How do ICQ theories fit in
this picture? An ICQ theory such as the one proposed in~\cite{eu} has $W$ bosons
identical to the SM's so the partonic interactions would not be different. But
once again ICQ and FCQ predictions are found to be the same and the explanation
is essentially the same used in single-photon processes: because no individual
colours are observed the result of any attempt to measure a quark's charge in 
this manner will be the colour average of the electric charge; for an up-type 
quark  the result of a jet-charge measurement should then be 
%%%%
\begin{equation}
Q^{jet}_{up} \;=\; \frac{1}{3}\; \sum_{i=1}^3\; q_{up}^i \;=\; \frac{1}{3}\; 
(1\,+\,1\,+\,0)\;=\; \frac{2}{3}\;\;\; ,
\end{equation} 
and likewise for a down-type quark, with the result $-1/3$. So, though 
ingenious, jet-charge measurements do not allow us to distinguish between ICQ
and FCQ models. 
\end{flushleft}

\begin{flushleft} $\bullet$ {\bf Meson radiative decays:} we 
already discussed dipolar transitions
in heavy mesons and argued they are predicted identically in both ICQ and FCQ 
models. There are however many experimental results of two-photon meson decays  
and these could in principle distinguish between both models. An important case 
is the decay of $\eta$ mesons into two photons, which have been used in the past
as evidence against ICQ theories~\cite{chan}. This is based on the fact that a
simple PCAC analysis of such decays yields for its width the
expression~\cite{berg}
\begin{equation}
\Gamma_{\gamma\gamma}^X \;=\; \left(\sum_{\mbox{colour}} \,<e_q^2>\right)^2\,
\frac{\alpha^2}{32\pi^3}\,\frac{m_X^3}{f_X^2} \;\;\;\ ,
\end{equation}
where $\alpha$ is the fine structure constant, $m_X$ and $f_X$ are the mass and
decay constant of the meson under consideration and $<e_q^2>$ the mean value of
the squared charges of the quarks in the state given by the meson's
wave function. At this point, in a simple approach, one considers the so-called
``nonet symmetry", that is, the equality of the decay constants for the mesons
belonging to the same flavour multiplet. The emblematic cases for comparison
between ICQ and FCQ theories are the pion, the $\eta_1$ and the $\eta_8$ states,
so we assume $f_1 = f_8 = f_\pi$. It is simple to show that for the pion and
$\eta_8$ states the widths are the same for both cases but, for the $\eta_1$,
the ICQ width is 4 times larger than the FCQ one. The first obvious retort to
this ``evidence" against ICQ theories is that it relies heavily on untested
theoretical assumptions, to wit the equality of the decay constants. There is in
fact no {\em a priori} reason why nonet symmetry should hold - in a na\"{\i}ve
quark model such an assumption might be plausible if the physical mesons $\eta$
and $\eta^\prime$ (which result from the mixing of the $\eta_1$ and $\eta_8$
states) were ideally mixed but as argued in ref.~\cite{chan} that is hardly the
case. Further, more elaborate calculations have found serious deviations from
nonet symmetry - for instance, using a Hidden Local Symmetry model, the authors
of ref.~\cite{ban} have found $f_1 \,=\,1.4 \, f_8$. Chanowitz~\cite{chan}
deduced equations for a $\xi$ parameter ($\xi = 1$ for FCQ, $\xi = 2$ for ICQ)
in terms of experimentally measured quantities which were in principle
independent of the nonet symmetry hypothesis and seemed to favour FCQ over ICQ.
The problem there is that from the start those equations were less reliable if
the theory being tested was ICQ. And it was shown later that even for the
``normal" FCQ theory Chanowitz's equations could not be applied in a na\"{\i}ve
manner as they did not take into account the existence of a more complex mixing
between $\eta_1$ and $\eta_8$~\cite{ben}. Pseudoscalar meson decays are an 
exceedingly difficult field of study, where great doubts still persist. For 
instance, different groups claim evidence for the presence of a significant 
gluon component in the $\eta$ and $\eta^\prime$ states~\cite{ball} or against 
it~\cite{ben}.
\end{flushleft}

Heavier quarkonia two-photon decays have also been studied in detail but one 
finds a situation similar to the lighter mesons - the FCQ prediction 
is~\cite{nov},
\begin{equation}
\frac{\Gamma (\eta_c \,\rightarrow\,\gamma \gamma)}{\Gamma (J/\Psi \,\rightarrow
\,e^+\,e^-)} \,=\,\,3\,Q_c^2 \;\frac{|\eta_c(0)|^2}{|J/\Psi(0)|^2}\;\;\; .
\label{eq:ch}
\end{equation}
If one {\em assumes} the equality of the meson's wave functions at the origin
(the same hypothesis that nonet symmetry is based on) the FCQ prediction
is a factor of $4/3$, whereas the ICQ result would be 3. The current
experimental values favour the FCQ result. However, to obtain this ``prediction"
we needed to make an extra assumption regarding the (unknown) wave functions
of quark bound states, and such an assumption is an oversimplification:
it has been shown~\cite{wong} that the $J/\Psi\,\rightarrow\,\gamma\,\eta_c$
magnetic transition is substantially underestimated if one assumes both
particles' wave functions are identical. Theoretical calculations of the $\eta_c$
two-photon amplitude also vary considerably (from 5 to 10 KeV~\cite{wong}, the 
experimental value being $7.2 \pm 1.2$ KeV, depending on whether one uses QCD 
sum rules~\cite{sumr} or a Bethe-Salpeter formalism~\cite{beth}). In fact, the 
authors of ref.~\cite{beth} find considerable differences between both mesons'
wave functions. With the charm quark mass dependence taken into account in
corrections to eq.~\eqref{eq:ch}, we find that if the $\eta_c$ and $J/\Psi$
wave functions differ by about only 20\%, the current experimental results would
actually favour ICQ theories. Like in the lighter mesons, then, the theoretical
description of the mesons' structure is much more complex than simple formulae
like~\eqref{eq:ch} lead to believe. With this in mind we must ask ourselves, are
we testing the character of the quarks's electric charges or our models about
their bound states? We argue that this uncertainty leaves room for speculation 
on ICQ models.

\begin{flushleft} $\bullet$ {\bf Final state radiation in hadronic 
interactions:} Brodsky and collaborators~\cite{bro} suggested an interesting 
method for establishing the character of the quarks' charges, by measuring the 
asymmetry between deep inelastic scattering of the proton using electrons or 
positrons. The difference between the cross sections for the processes $e^-\,p
\rightarrow e^-\,\gamma\,X$ and $e^+\,p\rightarrow e^+\,\gamma\,X$ would 
select the Bethe-Heitler-Compton interference term, proportional to the cube of 
the partons' electric charges at tree-level. This would in principle, then, 
provide us with a clear measurement of the values of the quarks' charges. The 
initial experiments~\cite{fan} were inconclusive, with results that could even 
be seen favoring the ICQ models. Further experiments were conducted, adapting 
this idea to final state radiation measured in $e^+\,e^- \rightarrow \mbox{2 
jets}$~\cite{fins}, and still, the results were inconclusive. A detailed 
theoretical analysis of these results was given in refs.~\cite{raj}, where it
was shown that the proposed measurements were not sensitive to the model 
considered. 
\end{flushleft}

The best argument against ICQ theories is an indirect one: the fact they 
(usually) predict gluons having mass and electric charge. The authors of 
refs.~\cite{pati,glu}, for instance, 
choose a gauge symmetry breaking mechanism that gives mass to all gluons all the
while preserving a global $SU(3)$ symmetry. Others have given mass to gluons by 
means of a four-vertex ghost field~\cite{fad}. Cornwall proposed a mechanism for
dynamic generation of gluon masses within a theory with unbroken colour gauge 
symmetry~\cite{corn}. According to the Particle Data Group~\cite{pdg} the 
current limit on the mass of the gluon is ``a few MeV". That, however, is a 
theoretical bound, derived from arguments that assume all gluons are degenerate
in mass~\cite{ynd}. Also, as others have discussed~\cite{fiel}, the arguments 
of~\cite{ynd} neglect to take into account the quantum field theoretical aspects
of the gluon-quark interactions, which casts doubt over their final conclusions.
The best experimental evidence for the existence of eight massless gluons is
the precision measurements agreeing with two and three loop QCD calculations, 
from which it is established that the running of the strong coupling constant is
in agreement with an $SU(3)$ $\beta$-function~\cite{pdg}, on an energy range 
from $\sim 2$ to 200 GeV. This remarkable result has to be taken into 
consideration when considering ICQ theories, whatever model we consider has to
reproduce the low-energy behaviour of $\alpha_S$. A trivial remark is that these
experimental results do not, in principle, contradict the existence of light 
massive gluons, with masses of the order of 1 GeV (possibly even larger, 
considering the experimental uncertainty in the values of $\alpha_S$ at low 
energies). This because the coefficients of  the $\alpha_S$ $\beta$-function at 
a given energy scale $M$ depend on the number of particles (both fermions and 
gluons) with masses smaller than $M/2$ - a crude explanation for this well-known
fact is to say that those are the particles ``circling" in the loops that 
contribute to the renormalisation of the gluonic propagator. Particles with larger
mass have been ``integrated out" of the theory (a simple example of this is the
change in the coefficient of $\beta_{\alpha_S}$ once the bottom quark 
threshold is crossed). It seems reasonable, then, that the $\alpha_S$ low-energy
evolution does not preclude ICQ models such as those considered in 
refs.~\cite{god,fiel}, with gluon masses smaller than $\sim 1$ GeV. In fact, in
ref.~\cite{fiel}, Field gives additional experimental arguments for the 
existence of massive gluons by showing that the decays of the $J/\Psi$ and 
$\Upsilon$ mesons are better explained assuming a gluonic mass of about 1 GeV. 
Furthermore, we must remark that it is not a necessary implication that ICQ 
models have no massless gluons - in ref.~\cite{eu} a model is constructed in 
which three gluons remains massless and neutral and five others gain mass, four
of which are charged. This example shows that quarks having integer charge is 
not necessarily tantamount to gluons having mass and charge. Again, we would
argue that these considerations - though hardly conclusive - leave us enough
room to speculate on the possibility of ICQ models. 

\section{Concluding remarks}

Any conclusions on this matter must necessarily be modest. The Standard Model, 
and
QCD in particular, are established theories responsible for a wealth of 
extremely precise predictions, well confirmed by numerous experiments. The
results from LEP2, however, are undeniable: at high energies, when our
perturbative models should work best, they fail to reproduce the experimental
results in four separate measurements. That these huge discrepancies occur in the 
two-photon channel is also very important - this is a direct probe into the
electric charges of quarks and the exact place where one would expect sizeable
differences between ICQ and FCQ models. They are a clear indication that 
something is wrong in our understanding of particle physics at very high 
energies, and we studied a particular solution for this problem. We have 
performed tree-level calculations using integer charged quarks and showed that 
such models reproduce the open production of heavy quarks in photon-photon 
collisions at LEP2. An estimation of higher order contributions to these cross 
sections was also made and indicates the agreement should remain if loop 
calculations are performed. ICQ theories also improve considerably the agreement
between theory and experiments on the inclusive production of pions and charged 
kaons via the two-photon channel at LEP2. Perhaps we could discount these 
results as numeric coincidence if we had found agreement with the data on a 
single observable, but in this paper we studied four different channels. Very 
good agreement with the data is even obtained for $\pi^0$'s and $K^\pm$'s by 
choosing the lowest fragmentation scale possible. We must remember that ICQ 
models improve the agreement with data for high $p_T$ values but we do not know 
what will happen in the low energy region once non-perturbative contributions 
are taken into account. We have reviewed the evidence for fractional quark 
charges and argued that it was not conclusive - the best arguments stem 
from two-photon meson decays but their analysis only points to FCQ under the
assumption of stringent theoretical assumptions. Nonet symmetry has proven to
be false and, for heavy mesons, the equality of wave functions is not supported
by relativistic calculations. The issue of gluon masses is a fundamental one, 
but there are ICQ theories that have massless gluons, and others for which the 
gluonic masses are so small as to not challenge, for instance, the low energy 
evolution of the strong coupling constant. Evidence for gluons being massless 
is also indirect, whereas the two-photon results provide direct evidence on
the quark charges. It is also worth mentioning that though ICQ models help 
explain the two-photon discrepancies found at LEP2, they will not, in principle,
have an impact on the excess of open bottom quark production observed at 
HERA~\cite{hera,rev}. Regardless of whether one believes in ICQ models or not, 
it seems clear they do a better job than the SM at describing the two-photon 
data. As such they are, to the best of our knowledge, at present the only 
explanation for the discrepancies observed at LEP2. We cannot and should not
throw away the Standard Model on these experimental disagreements alone - but if
as time passes we find ourselves unable to find an explanation for them in the 
framework of the SM new ideas will have to be considered. We believe it is 
worthwhile to make the observation that ICQ-type theories are a possible
way out of this problem. If indeed no other alternative presents itself ICQ 
theories may have to be reconsidered. At this point more refined
calculations, involving higher order contributions and those coming from 
resolved processes, are necessary to make sure that the agreement with the data 
we found here remains. It would also be desirable to obtain data on inclusive
hadron production in the two-photon channel from LEP collaborations other than 
L3 - despite their excellent work in studying these channels, independent 
confirmation of large discrepancies with the SM would clearly be reassuring. 

\vspace{0.25cm}
{\bf Acknowledgments:} I thank Pablo Achard, Augusto Barroso, Mario Greco and Rui 
Santos for discussions and suggestions. This work was supported by a fellowship from 
Funda\c{c}\~ao para a Ci\^encia e Tecnologia, SFRH/BPD/5575/2001.

\end{document}